\documentclass[a4paper,11pt]{article}
\usepackage{pos}
\usepackage{booktabs}
\usepackage{natbib}

\title{Gravitational--wave cosmology with dark sirens: state of the art and perspectives for 3G detectors}
\ShortTitle{Gravitational-wave cosmology with dark sirens}

\author*[a,b,c,d]{Michele Mancarella}
\author[f,g]{Nicola Borghi}
\author[c,d]{Stefano Foffa}
\author[e]{Edwin Genoud-Prachex}
\author[c,d]{Francesco Iacovelli}
\author[c,d]{Michele Maggiore}
\author[f,g]{Michele Moresco}
\author[f]{Matteo Schulz}

\affiliation[a]{Dipartimento di Fisica ``G. Occhialini'', Universit\'a degli Studi di Milano-Bicocca,
 Piazza della Scienza 3, 20126 Milano, Italy}

\affiliation[b]{INFN, Sezione di Milano-Bicocca, 
Piazza della Scienza 3, 20126 Milano, Italy}

\affiliation[c]{D\'epartement de Physique Th\'eorique, Universit\'e de Gen\`eve, 
24 quai Ansermet, CH-1211 Gen\`eve 4, Switzerland}

\affiliation[d]{Gravitational Wave Science Center (GWSC), Universit\'e de Gen\`eve,  CH-1211 Gen\`eve 4, Switzerland}

\affiliation[e]{Institute for Theoretical Physics, Goethe University, 
60438 Frankfurt am Main, Germany}

\affiliation[f]{Dipartimento di Fisica e Astronomia ``Augusto Righi'', Alma Mater Studiorum Universit\`a di Bologna, 
via Piero Gobetti 93/2, I-40129, Bologna, Italy}

\affiliation[g]{INAF - Osservatorio di Astrofisica e Scienza dello Spazio di Bologna, 
via Piero Gobetti 93/2, I-40129, Bologna, Italy}

\emailAdd{michele.mancarella@unimib.it}
\emailAdd{nicola.borghi6@unibo.it}
\emailAdd{stefano.foffa@unige.ch}
\emailAdd{genoud@itp.uni-frankfurt.de}
\emailAdd{francesco.icovelli@unige.ch}
\emailAdd{michele.maggiore@unige.ch}
\emailAdd{michele.moresco@unibo.it}
\emailAdd{matteo.schulz@studio.unibo.it}

\abstract{ A joint fit of the mass and redshift distributions of the population of Binary Black Holes detected with Gravitational--Wave observations can be used to obtain constraints on the Hubble parameter and on deviations from General Relativity in the propagation of Gravitational Waves. 
We first present applications of this technique to the latest catalog of Gravitational--Wave events, focusing on the comparison of different parametrizations for the source--frame mass distribution of Black Hole Binaries. We find that models with more than one feature are favourite by the data, as suggested by population studies, even when varying the cosmology.
Then, we discuss perspectives for the use of this technique with third generation Gravitational--Wave detectors, exploiting the recently developed Fisher information matrix \texttt{Python} code \texttt{GWFAST}.}

\FullConference{%
  41st International Conference on High Energy physics - ICHEP2022\\
  6-13 July, 2022\\
  Bologna, Italy
}

\begin{document}
\maketitle

\section{Introduction}

Gravitational Waves (GWs) from compact binaries are 
a new player in cosmology, following their first direct detections in recent years. 
In particular, GWs provide a new way to test the luminosity distance--redshift relation, based on the fact that they give a direct measurement of the luminosity distance to the source. At low redshift, this can be used to obtain a measurement of the Hubble constant $H_0$ \cite{Schutz:1986gp}. At higher redshift, the luminosity distance as measured by GWs can carry the imprint of deviations from General Relativity (GR) in the form of extra friction experienced by GWs during their propagation, a phenomenon known as ``modified GW propagation'' \cite{Belgacem:2017ihm}. The amplitude of this effect is encoded in an extra parameter $\Xi_0$ in the luminosity distance--redshift relation, with GR defined by $\Xi_0=1$ \cite{Belgacem:2018lbp}. Viable models predict values as large as $\Xi_0=1.8$ \cite{Belgacem:2019lwx}.

Applications of different techniques to measure both $H_0$ and $\Xi_0$ with GWs have started in recent years \cite{Abbott:2017xzu, Abbott:2019yzh, Ezquiaga:2021ayr, Finke:2021aom, LIGOScientific:2021aug, Palmese:2021mjm, Mancarella:2021ecn, Leyde:2022orh}.
Due to the perfect degeneracy between mass and redshift in the GW waveform of Binary Black Hole systems, the latter cannot be directly determined with observations of GWs from these sources. The detection of an electromagnetic (EM) counterpart can give an independent measurement of the redshift, but this has been the case so far only for the Binary Neutron Star GW170817, which led to a measurement of the Hubble constant \cite{Abbott:2017xzu}, while being at too low redshift to allow a good determination of the parameter $\Xi_0$, which was bound only at the level $\Xi_0 \lesssim \mathcal{O}(10)$ \cite{Belgacem:2018lbp}. In absence of a counterpart, the source goes under the name of ``dark siren'' and information on the redshift can be obtained at the statistical level with an analysis of the full population of sources. This allows to exploit all the GW events, while being of course subject to larger errors that should be compensated by the size of the catalog.
The first goal of this contribution is to present measurements of both $H_0$ and $\Xi_0$ from dark sirens with the use of information from the mass distribution of Binary Black Holes (BBHs) obtained with the latest GW data releases \cite{LIGOScientific:2021djp} and with the code \texttt{MGCosmoPop} \raisebox{-1pt}{\href{https://github.com/CosmoStatGW/MGCosmoPop}{\includegraphics[width=10pt]{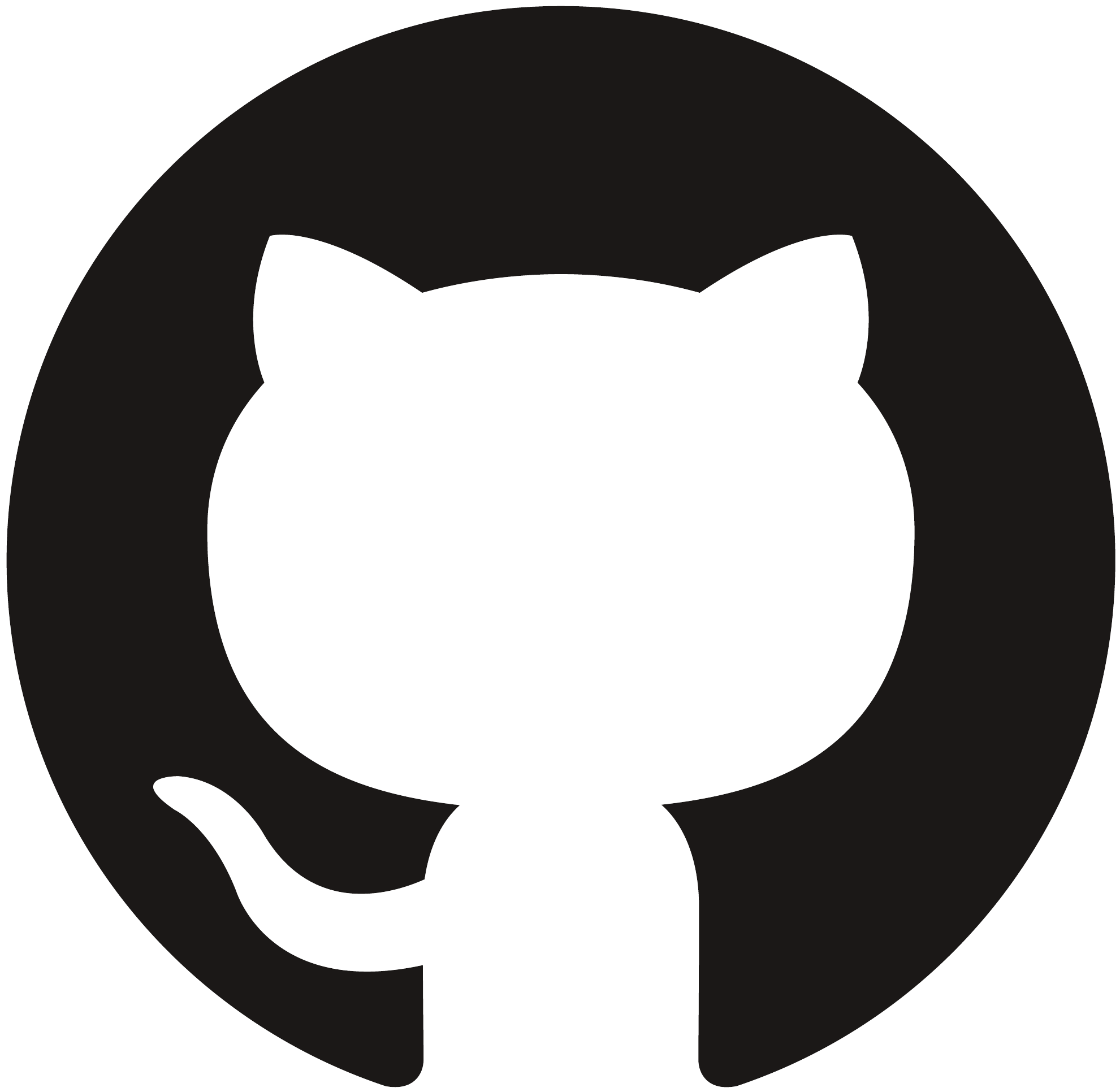}}}~\footnote{\url{https://github.com/CosmoStatGW/MGCosmoPop}} \cite{Mancarella:2021ecn}, comparing results obtained with different assumptions for the BBH mass distribution.

Even if such results are a remarkable progress for cosmology, the constraining power of the measurements remains far from the one already achieved by EM observations on the $\Lambda$CDM parameters.
The reason mainly resides in the limited size of the available GW catalogs and limited sensitivity of current second--generation (2G) GW detectors.
The situation will drastically change with third--generation (3G) ground--based GW detectors currently under design, i.e. the Einstein Telescope in Europe \cite{Punturo:2010zz, Maggiore:2019uih} and the Cosmic Explorer in the US \cite{Evans:2021gyd}. These will undergo an improvement of at least one order of magnitude in sensitivity, while substantially broadening the frequency range of the instruments towards lower frequencies. Such improvements will lead to precision measurements of source parameters for a huge number of sources, opening the possibility of precision measurements in cosmology with GWs.
A crucial point for characterizing science with 3G detectors is to forecast their capability of measuring the source parameters for population of sources calibrated on the latest constraints from current GW detectors. In the second part of this contribution, we will present \texttt{GWFAST} \raisebox{-1pt}{\href{https://github.com/CosmoStatGW/gwfast}{\includegraphics[width=10pt]{GitHub-Mark.pdf}}}~\footnote{\url{https://github.com/CosmoStatGW/gwfast}} \cite{Iacovelli:2022mbg}, a new Fisher--information matrix \texttt{Python} code to rapidly forecast the detection capabilities of networks of 3G detectors. We will then discuss perspectives for the techniques applied to the latest GW data releases, based on forecasts obtained with this new code.

\section{Constraints on the Hubble constant and modified gravitational--wave propagation with GWTC--3}\label{sec:2G}
\begin{figure}[h]
    \centering
    \begin{tabular}{c@{\hskip 3mm}c}
  \includegraphics[width=64mm]{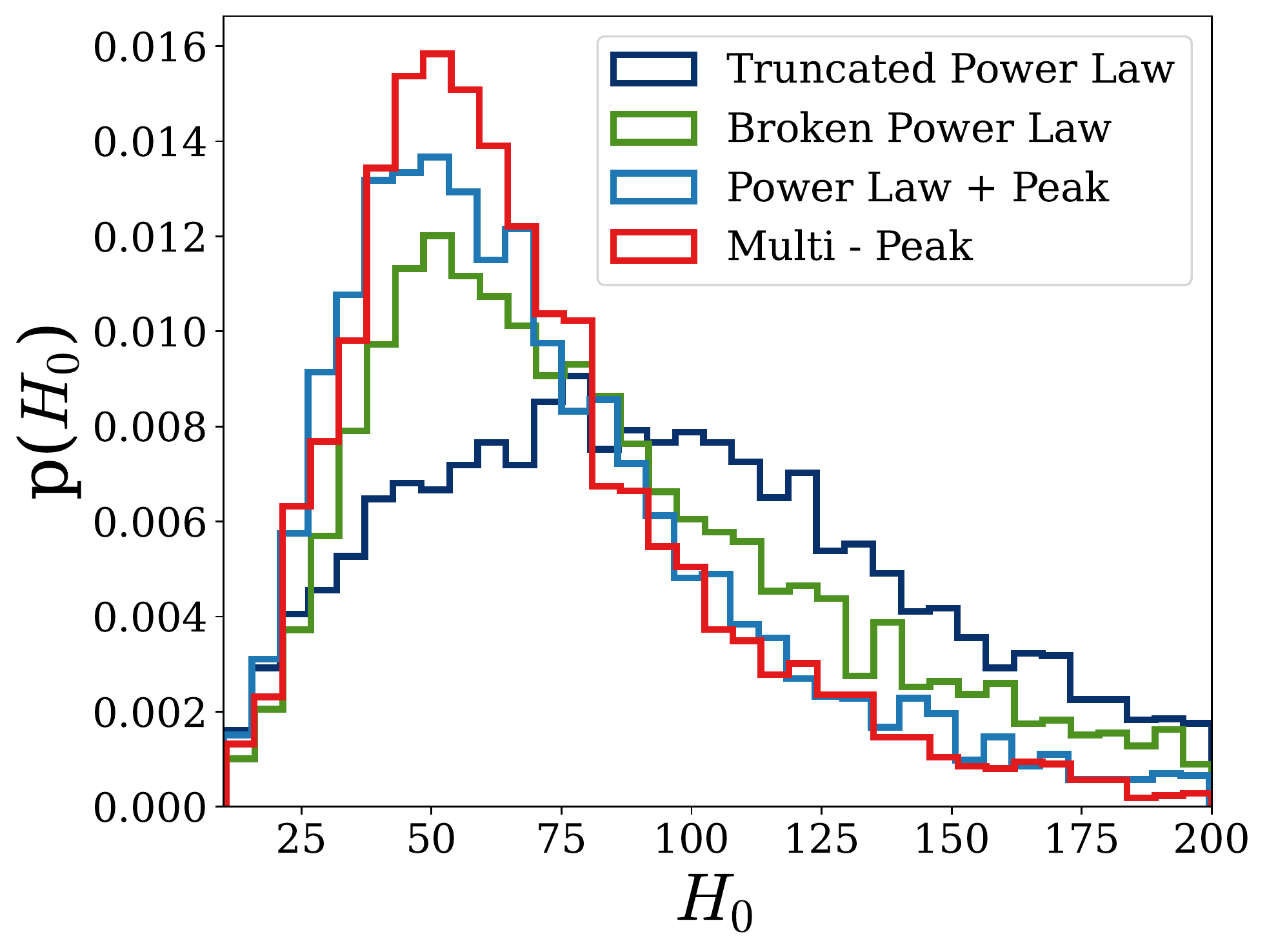} &   \includegraphics[width=64mm]{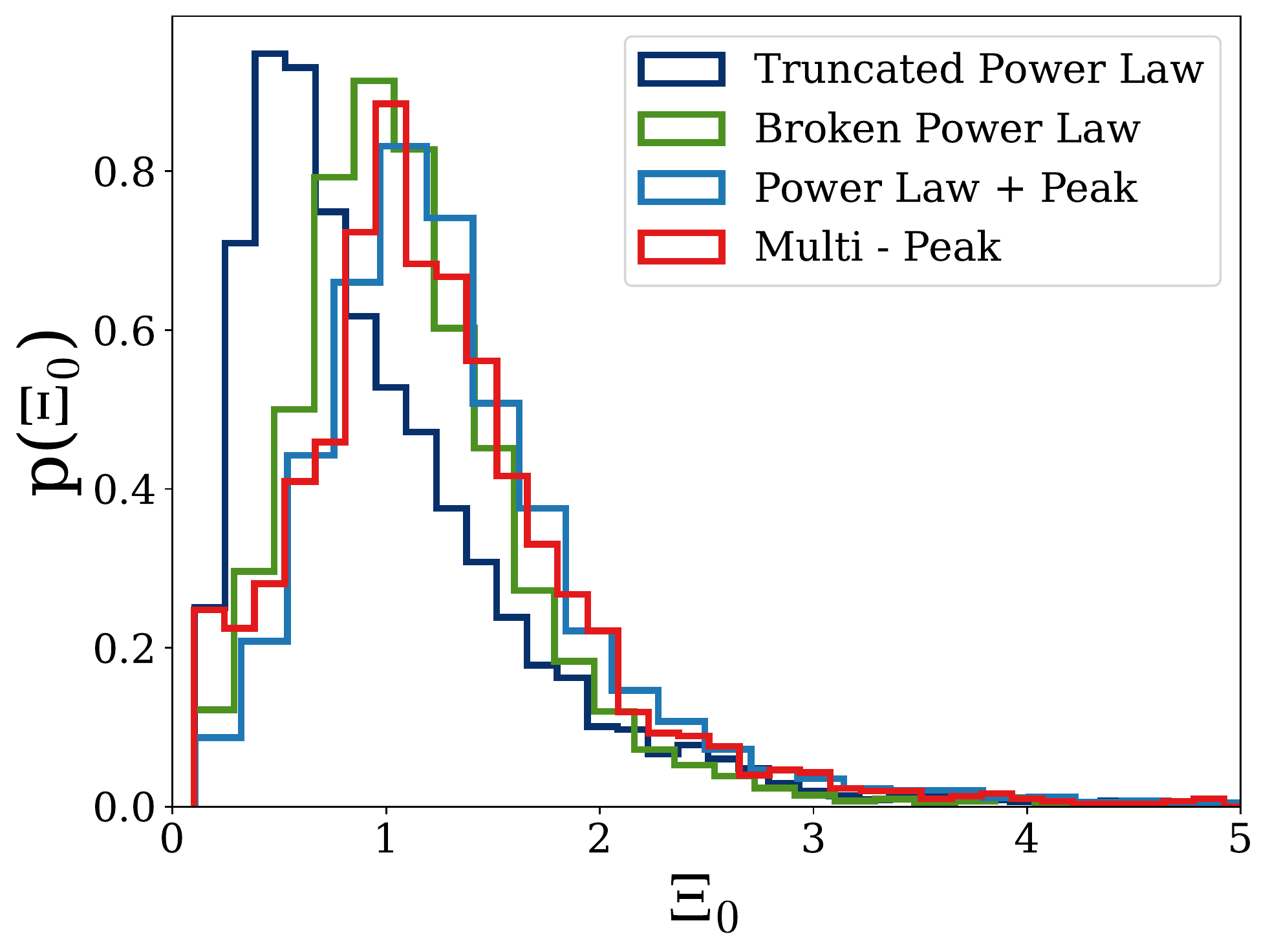}
\end{tabular}
    \caption{Left: marginal posterior probability distribution for $H_0$ in case (i), i.e. assuming a flat $\Lambda$CDM universe ($\Xi_0=1$). Right: marginal posterior probability distribution for $\Xi_0$ in case (ii), i.e. assuming a Planck 2018 background cosmology. Different colors correspond to different models for the BBH mass distributions.}
    \label{fig:massFunction}
\end{figure}
\begin{table}[h]
    \centering
    \begin{tabular}{ |c||c|c||c|c|c| }
    \toprule\midrule
     Mass Model & case (i): $H_0$  & DIC for case (i) & case (ii): $\Xi_0$ &  DIC for case (ii)\\
    \midrule\midrule
     Truncated Power Law & $91_{-64}^{+84}$  & -18.29 & $0.8_{-0.5}^{+1.5}$ &  -13.93 \\
    \midrule
    Broken Power Law & $74_{-45}^{+91}$  & -7.66 & $1.0_{-0.7}^{+1.1}$ &  -5.02 \\
   \midrule
    Power Law + Peak & $63_{-37}^{+83}$ & -3.59 & $1.2_{-0.7}^{+1.4}$ & -0.11  \\
   \midrule
    Multi -- Peak & $61_{-35}^{+74}$   & 0 & $1.2_{-0.9}^{+1.4}$ & 0  \\
    \midrule\bottomrule
    \end{tabular}
    \caption{ Median and $90\%$ symmetric C.I. for case (i), i.e. $H_0$ assuming a flat $\Lambda$CDM universe ($\Xi_0=1$), and case (ii), i.e. $\Xi_0$ assuming a Planck 2018 background cosmology, for the four models of the BBH mass distribution. For each mass model (and for both $H_0$ and $\Xi_0$), we also report the difference between the Deviance Information Criterion (DIC) of the ``multi--peak'' model, which is favoured in both cases, and each other model. Larger negative values correspond to more disfavoured models. }
    \label{tab:constraints}
\end{table}
One possibility to break the mass--redshift degeneracy of the GW waveform is to exploit the presence of a feature in the BBH mass distribution, predicted by the theory of Pair Instability Supernova \cite{Spera:2017fyx} to be around $\sim 30-50\ M_{\odot}$. With a hierarchical Bayesian analysis which correctly accounts for selection bias \cite{Mandel:2018mve}, it is possible to  obtain a measurement of the cosmological parameters, marginalised over the parameters of the distributions of masses and redshift of the population \cite{Farr:2019twy}. This technique has been applied to $H_0$ \cite{LIGOScientific:2021aug, Mancarella:2021ecn} and to modified GW propagation \cite{Ezquiaga:2021ayr,Mancarella:2021ecn,Leyde:2022orh}. 
In this contribution we consider four models for the mass distribution: a ``truncated power law'', a ``broken power law'', a ``power law + peak'' and a power law with two Gaussian peaks (denoted as ``multi--peak''). For the definition of these models we refer to Appendix B of \cite{LIGOScientific:2020kqk}. In population analyses with fixed cosmology, models with more than one feature seem to be preferred, showing in particular the possible presence of a secondary peak around $\sim 10\ M_{\odot}$ \cite{LIGOScientific:2021psn}. It is therefore of interest to check if this conclusion is robust to a variation of the cosmology, and in turn if the presence of extra features impacts the cosmological constraints.
For each of the four mass models, we consider separately two cases: (i) $\Xi_0$ is fixed to the GR value of 1, and $H_0$ is inferred; and (ii) the background expansion history is fixed by Planck 2018 (i.e. $H_0=67.66\ \rm Km/s/Mpc$, see \cite{Planck:2018vyg}), and the modified GW propagation parameter $\Xi_0$ is inferred. We refer to \cite{Mancarella:2021ecn} for further details on the methodology and the choice of the sample used \footnote{We remind here that we select the GWTC-3 events with signal--to--noise ratio (SNR) larger than 12. }.
Results are obtained with the code \texttt{MGCosmoPop} \raisebox{-1pt}{\href{https://github.com/CosmoStatGW/MGCosmoPop}{\includegraphics[width=10pt]{GitHub-Mark.pdf}}} \cite{Mancarella:2021ecn}.

Figure \ref{fig:massFunction} presents the marginal posteriors for case (i) (left) and (ii) (right). Different colors refer to results obtained using each of the four mass models described above.
The values of the median and $90\%$ symmetric Confidence Interval (C.I.) for $H_0$ and $\Xi_0$ are reported in Table \ref{tab:constraints}, together with a comparison of the goodness of fit for the four models, using the deviance information criterion (DIC) \cite{Spiegelhalter:2002yvw}, defined as $\rm{DIC} = -2 \left[ \langle{\log {\mathcal{L}}}\rangle-\rm{var}[\log{\mathcal{L}}] \right]$, normalized to the best performing model. Negative values indicate that the model is disfavoured.
Consistently with other works \cite{Leyde:2022orh}, we find that the ``multi--peak'' model is favoured by data. Interestingly, this preference for more complex models is in line with the results obtained from population studies with a fixed cosmology \cite{LIGOScientific:2021psn}, even if letting the cosmological parameters vary inevitably broadens the constraints on the parameters of the mass distribution. The best--fit model gives a $\sim 90\%$ constraint on $H_0$ and a $\sim 95\%$ constraint on $\Xi_0$ ($90\%$ C.I.). These error are quite large and limited by the small sample size and on the statistical error on individual source parameters. However, it is worth noticing that the constraint on the parameter $\Xi_0$ represents the tightest one currently available from GW measurements. 
\section{Perspectives for third--generation gravitational--wave detectors}
An intense activity to develop a third generation (3G) of GW detectors is ongoing in the current period. In particular, the Einstein Telescope in Europe has been included in the roadmap of large European scientific infrastructures \footnote{See \url{https://www.esfri.eu/latest-esfri-news/new-ris-roadmap-2021.}}. A crucial aspect for describing science at 3G detectors is to be able to compute efficiently the signal--to--noise ratio (SNR) and parameter estimation capabilities of networks of detectors. For this purpose, a number of tools bases on the Fisher information matrix formalism \cite{Vallisneri:2007ev} have been developed recently, in particular \texttt{GWBENCH} \cite{Borhanian:2020ypi}, \texttt{GWFISH} \cite{Harms:2022ymm}, and \texttt{GWFAST} \cite{Iacovelli:2022mbg} (see also \cite{Li:2021mbo, Pieroni:2022bbh}). These codes have undergone an accurate cross--validation process \cite{Iacovelli:2022bbs}. 

\begin{figure}[t]
    \hspace{-2cm}
    \begin{tabular}{c@{\hskip 3mm}c@{\hskip 3mm}c}
  \includegraphics[width=60mm]{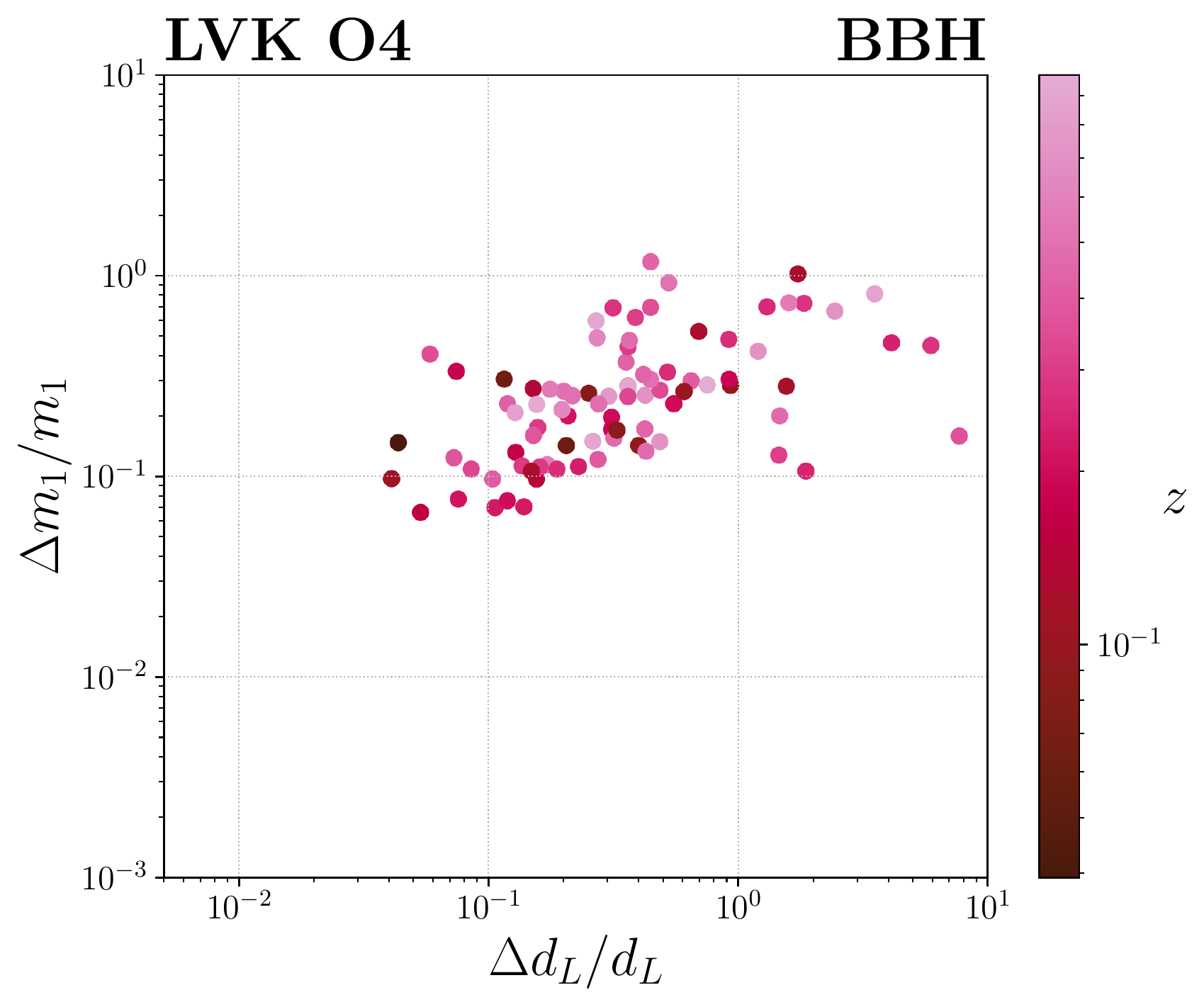} &   \includegraphics[width=60mm]{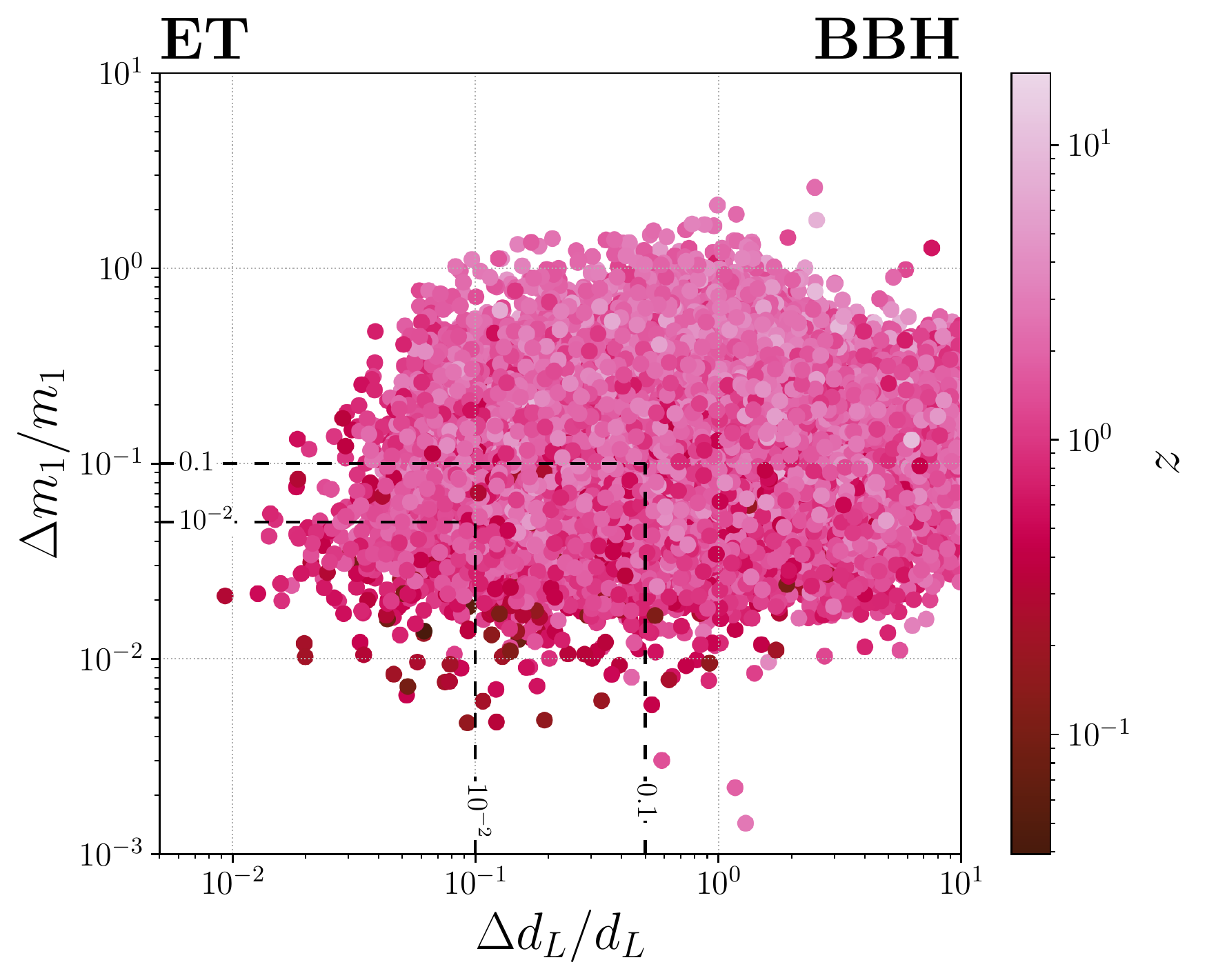} &
  \includegraphics[width=60mm]{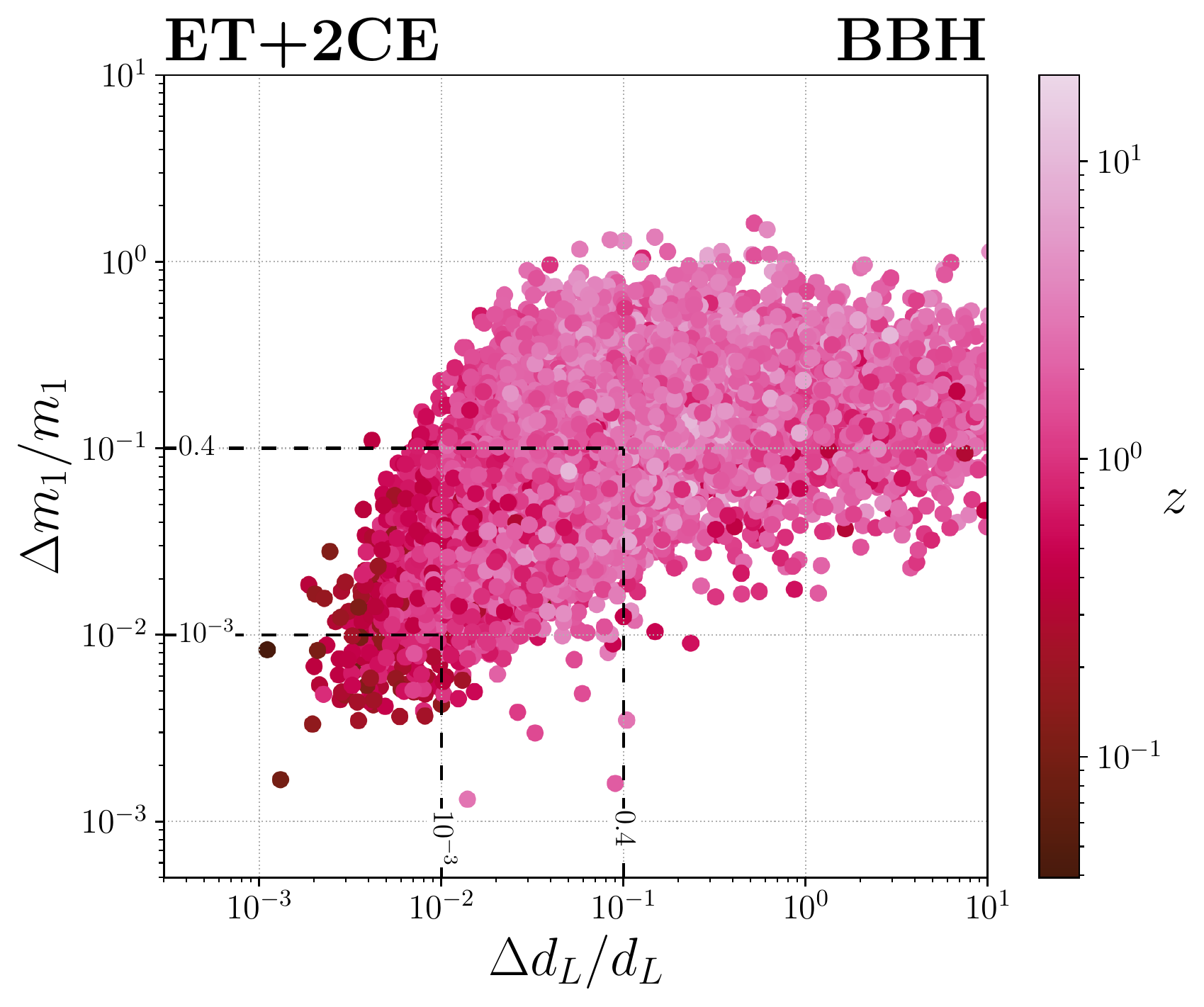}
\end{tabular}
    \caption{Forecasted relative errors on the luminosity distance and detector--frame masses for the O4 run of the LVK collaboration (left), for an ET detector (central), and for a network of ET and two CE detectors (right), during one year of observations. Numbers on the dashed lines indicate the fraction of sources enclosed in the corresponding region.}
    \label{fig:forecast}
    \vspace{-.2cm}
\end{figure}

In particular, \texttt{GWFAST} \raisebox{-1pt}{\href{https://github.com/CosmoStatGW/gwfast}{\includegraphics[width=10pt]{GitHub-Mark.pdf}}} \cite{Iacovelli:2022mbg} is a \texttt{Python} code based on Automatic Differentiation \cite{https://doi.org/10.1002/widm.1305}, which makes use of the library \texttt{JAX} \cite{jax2018github}. This allows efficient parallelization and numerical accuracy.
As an application of \texttt{GWFAST}, we compute the predicted distribution of relative errors on the parameters that are more relevant for the application of the dark siren techniques discussed in Sec. \ref{sec:2G}, namely the detector--frame primary mass $m_1$ and luminosity distance $d_L$, as measured by the LIGO--Virgo--Kagra (LVK) network during the upcoming O4 observing run, by an ET detector and by a network of ET and two Cosmic Explorer (CE) detectors, in all cases during one year of observations. The simulated population is tuned on the latest observations of the LVK \cite{LIGOScientific:2021psn}. We refer to \cite{Iacovelli:2022bbs} for details on the simulation.
Figure \ref{fig:forecast} presents scatter plots for the forecasted statistical error in the plane $(\Delta d_L /d_L - \Delta m_1 / m_1)$. 
The quantitative improvement of 3G detectors is clear from a visual inspection of the figure, both for the number of detected sources ($86$ for O4 against $\sim 5.1 \times 10^{4}$ for ET and $\sim 7.0 \times 10^{4}$ for ET+2CE) and for the accuracy of the measurements. In particular, as anticipated in the introduction, this result shows that percent--level measurements on the source parameters will be possible for a large number of sources, opening the possibility of obtaining precision results with dark siren techniques such as the one applied in Sec. \ref{sec:2G}, which with the current generation of detectors remain limited by the rather large statistical errors on masses and distances that are obtained for the majority of the events.
Therefore, the transition to 3G GW detectors will open the possibility of precision GW cosmology with dark sirens.

\bibliographystyle{utphys}
\bibliography{myrefs}

\end{document}